# CLIC Detector Main Solenoid Design & Status Report

Andrea Gaddi[1], Benoit Curé[1] and Alain Hervé[2]

1 – CERN – Physics Department
Geneva, Switzerland

2 – University of Wisconsin
Wisconsin – USA

The magnet system for the CLIC Detector concepts is composed of the central solenoid in combination with the two forward anti-solenoids and the ancillary systems necessary for their operation, including the so-called push-pull scenario, allowing the quick exchange of the two detectors on the beam line. An overview of the design parameters of the detector main solenoids is presented hereafter.

## 1 Introduction

The conceptual design of the central solenoid is based on the experience gained in the past fifteen years with the construction and operation of the LHC detector magnets, in particular the ATLAS central solenoid (Ref. 1) and the CMS solenoid (Refs. 2,3,4). This experience is complemented with the results of conceptual design studies performed for the ILC detector magnets (Refs. 5,6) as well as of R&D campaigns and design studies for high field thin solenoids made at KEK (Ref. 7) and at CERN (Ref. 8).

## 2 Magnetic field design

The main parameters of the superconducting central solenoids of the two detectors are listed in Table 1.

|  | Nominal magnetic field (T) | Free bore (mm) | Magnetic length (mm) | Cold mass weight (tons) |
|---|---|---|---|---|
| CLIC_SiD | 5.0 | 5480 | 6230 | 170 |
| CLIC_ILD | 4.0 | 6850 | 7890 | 210 |

The nominal magnetic field is the value on the coil axis at the interaction point. The free bore corresponds to the inner diameter of the coil cryostat and the active length to the length of solenoid windings. The cold mass weight includes coil windings and the outer support cylinder. Magnetic field calculations have been performed to evaluate the magnetic field map in the detector tracking volume in order to estimate the magnetic field homogeneity, which is very stringent, especially for CLIC_ILD due to the choice of a TPC as central tracking device. The magnetic field distortion along the z-direction within the size limits (Z,R) of the TPC is defined as

$$\Delta l(r) = \int_0^Z \frac{B_r(r)}{B_z(r)} dz$$

and is required to be less than 10 mm everywhere. This requirement on the magnetic field quality can somehow be relaxed assuming that it is possible to precisely measure, at the level of better than $10^{-4}$ the actual magnetic field map once the magnet is built (Refs. 9,10). The precision of this measurement depends on the resolution of the magnetic field sensors and the accessibility of the volume where the magnetic field scan has to be carried out. Also at the time of the magnetic field mapping, the assembled detector has to be nearly in its final configuration in order to have the final magnetic environment present. In addition, the magnet power supply has to deliver the nominal current with a ripple of less than a few ppm. This looks feasible when considering the performance of the LHC detector magnet power supplies.

Figure 9.1 Magnetic field (T) generated by the central solenoid in CLIC_SiD.

The magnetic stray field outside the detector is calculated as well. This parameter is important because the proximity of the iron of the second adjacent detector can eventually cause perturbations in the magnetic field chart of the first detector. The magnetic stray field is reduced by choosing an appropriate thickness of the return yoke, in particular for the detector end caps. A conflicting argument, however, is the overall weight of the detector, which has to be limited in order to ease the assembly, facilitate the push-pull operation and keep costs under control. Consequently a certain residual stray field of about 5 mT at 15 m distance from the detector center has to be accepted.

The iron yoke is also important for radiation protection. The off-beam detector in its garage position is positioned adjacent to the operational on-beam detector. The personnel working on the parked detector is exposed to radiation escaping from the adjacent on-beam detector. A thick and nearly hermetically sealed iron yoke acts as a radiation shield and greatly helps to reduce the dose to personnel. Nevertheless, thick shielding doors are foreseen between the two detectors to be used when the actual dose is higher than expected or in case the legally tolerable dose is decreased by the time of operation.

## 3    Solenoid coil design & conductor options

### 3.1    Coil design

The main parameters of the central solenoid of the CLIC_SiD design are listed in Table 2 (Ref. 11). They are similar to those of the CMS solenoid. Thus the design relies on many features successfully tested before:
-   an aluminum stabilized superconductor with a mechanical reinforcement,
-   a multi-layer and multi-module coil with an external support cylinder,
-   a coil winding technique adapted for a fiber glass wrapping of the conductor,
-   vacuum impregnation of each module and a subsequent curing heat treatment,
-   an indirect, conduction based cooling by Helium flow in Al tubes attached to the surface of the support cylinder,
-   operation in a thermo-siphon cooling mode,

- a supporting system with radial and axial tie rods attached to the external mandrel and to the vacuum tank.

| Nominal magnetic field at the IP | 5.0 T |
|---|---|
| Peak magnetic field on the conductor | 5.8 T |
| Free bore diameter | 5.5 m |
| Magnetic length | 6.2 m |
| Ampere.turns | 34 MA.turns |
| Operating current | 18 kA |
| Stored magnetic energy | 2.3 GJ |
| Energy/Mass ratio | 14 kJ/kg |
| Inductance | 14 H |

Table 2. Main parameters of the superconducting central solenoid for CLIC_SiD.

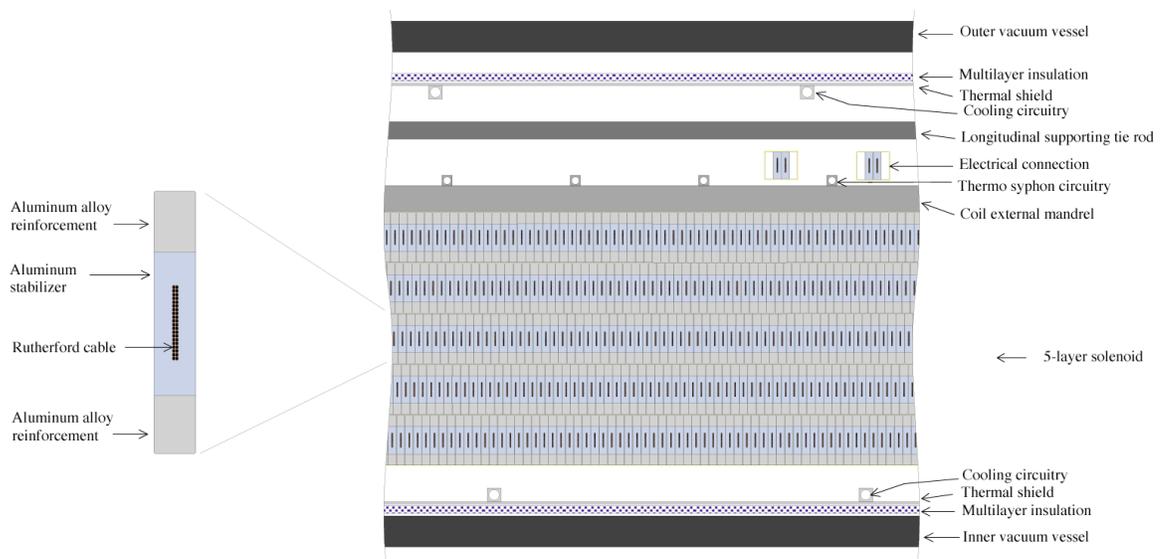

Figure 1. Sketch of the superconducting solenoid longitudinal cross section.

The number of layers in the coil windings is kept at a minimum while maintaining a realistic aspect ratio (height/width) of the conductor below 7. This leads to a 5-layer coil in combination with a conductor cross section of 97.4 mm x 15.6 mm. The total length of conductor is about 40 km. The solenoid is split in 3 coil modules and 5 layers per coil module means that the conductor unit length required is a fair 2.7 km. There are 16 layer-to-layer and module-to-module electrical connections. In order to have maximum temperature margin in these connections they are positioned in a region of low magnetic field on the outer radius of the mandrel, similarly to what as been done for CMS. The thickness of the support cylinder is limited for manufacturing reasons and a 50 mm thick mandrel is a reasonable choice.

The conductor design fulfills the requirements imposed for proper superconducting performance, thermal and electro-magnetic stability, mechanical strength, and quench protection.

A temperature margin of 1.5 K is considered realistic and possible when using a 40-strands NbTi/Cu Rutherford cable using state-of-the-art critical current density in NbTi conductors of 3000 A/mm$^2$ at 4.2 K and 5 T. At an operating current of 18 kA the solenoid operates at 32 % of the conductor critical current at an operating temperature of 4.5 K and the 5.8 T peak magnetic field in the solenoid innermost layer. The enthalpy margin under these conditions is 2.4 J/m related to the conductor cross-section. The 4.5 K operating temperature is compatible with static and transient heat loads. The radiation load is intercepted by the thermal radiation shield cooled at 80 K. The thermal shield is covered by multilayer super-insulation. The cryostat isolation vacuum of 10$^{-6}$ mbar at 4.5 K is a realistic target.

The mechanical reinforcement of the conductor can be realized by applying existing technologies, either by welding one or two structural aluminum alloy bars to the aluminum sheathed Rutherford cable (like the conductor for the CMS solenoid), or by using a micro-alloyed aluminum stabilizer hardened by a final cold working operation (as used for the conductor of the ATLAS central solenoid). A combination of both reinforcement technologies may be the best solution. Presently an R&D program is underway to develop such an advanced aluminum stabilized conductor. A trial production of a few prototype conductor units is launched to help in the decision-making process. The dimensions of the conductor and its reinforcement are defined to keep the maximum stress well below the tensile and yield strength of the reinforcing material. Safety margins of typically 1/3 of the tensile strength and 2/3 of the yield strength are respected. A detailed structural analysis will be carried out to confirm the estimations. The Residual Resistivity Ratio (RRR) of the stabilizer that remains at the end of the solenoid manufacturing and during the solenoid lifetime is kept at about 300 in order to satisfy the superconductor stability requirement.

The quench protection of the solenoid is ensured by the extraction of a significant amount of the stored magnetic energy and dump into an external resistor. The discharge rate is effective to use the so-called quench-back effect by which the ohmic heating by the current induced in the aluminum alloy support cylinder contributes significantly to cause a uniform warming up of the solenoid windings. The fast discharge voltage is kept below 600 V. The computed temperature gradient between two adjacent layers is less than 7 K, and about 60% of the stored energy is extracted. The average temperature of the solenoid cold mass after a fast dump stays below 70 K. In the ultimate fault case of just normal zone propagation and no external energy extraction, the quench-back effect is still efficient and the peak temperature will not exceed 150 K.

The support cylinder sections shall be carefully manufactured to fulfill the dimensional tolerances and to maintain the mechanical properties of the material. The technique used for CMS using reference tooling and machining in steps, in combination with seamless flanges obtained by ring rolling and their welding to the cylindrical mandrel, looks feasible for this design as well.

The module-to-module mechanical coupling is performed, as with CMS, by stacking the modules with their axis vertical while carefully controlling the flatness of the contact surfaces. In this way, the assembly of the solenoid can be done in a vertical arrangement without imposing cylindrical deflections that would complicate the modules coupling. Finally the solenoid is rotated by 90 degree to allow its insertion into the cryostat.

### 3.2 Conductor options

To provide the required reinforcement of the aluminum stabilized conductor, the baseline is to adapt the existing strengthening technologies that were applied with success on thin superconducting coils for particle detectors in high energy physics, and to identify and investigate the emerging technologies that may become industrially available at the time the project will be launched.

Currently, the options considered are, either single or in combination:
- add by welding structural bars from aluminum alloy (used in the CMS solenoid),
- use a Zn or Ni doped, micro-alloyed structural stabilizer (used in the ATLAS solenoid),
- identify and develop a novel doping method for stabilizer reinforcement, such as carbon nano-tubes, of which availability and effectiveness for an aluminum stabilized superconductor has still to be demonstrated.

In accordance with the present design, the conductor cross section is shown in Figure 2 together with the cross sections of several other aluminum stabilized conductors manufactured for recent large superconducting detector magnets.

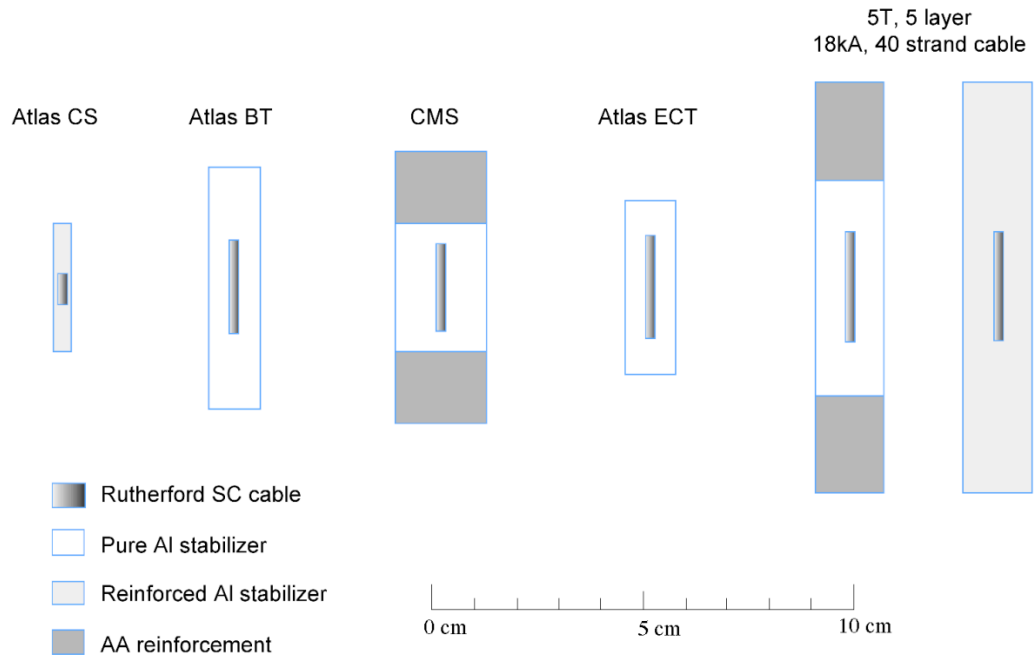

Figure 2. Cross sections of Al stabilized and reinforced conductors previously used and the proposed two conductor options for achieving 5 T in the CLIC_SiD design.

Since the feasibility of the conductor technology is crucial for the design, an R&D program is launched to develop the conductor options and demonstrate the feasibility of manufacturing the proposed large conductor made with a structural stabilizer. As a first trial a 100 m demonstration length will be co-extruded and cold worked using an Al-0.1wt%Ni, identical to the material used in the ATLAS solenoid. The manufacturing of the demonstration sample will allow the measurement of the mechanical properties, i.e. the gain in tensile and yield strength, the RRR of the conductor stabilizer, the uniformity of the properties across the very large section and the quality of the inter-metallic bonding between the Rutherford cable and the stabilizer. In addition the superconducting properties are investigated to check for possible degradation effects on the critical current density, determine induced ramp losses and verification of the normal zone propagation. Based on this experience, further steps in the conductor R&D program can be decided upon.

## 4   Magnet services & push-pull scenario

The detector magnets will be switched off during the push-pull operation, however they will be kept at cryogenic temperatures during the moving. It is required that the push-pull operation can be carried out in a short lapse of time. Therefore the magnet services (cryogenics and vacuum pipes, powering and protection lines) have to be compatible with constraints imposed by the push-pull scenario. Already on the ATLAS End Cap Toroids, flexible cryogenics transfer lines and vacuum pipelines are in use to allow the opening of the detector and provide access to the detector interior.

Considering the overall length of the powering line that could easily exceed 60m, and to avoid repetitive opening and closing of the circuit via bolted connections during the detector push-pull, with the associated risk of resistance increase and consequently localized overheating, a permanent flexible connection is considered. An option is a high temperature superconducting flexible cable assembly (Ref. 12) cooled by the helium gas return line at a temperature between 5 and 20 K. A second option would be a NbTi based flexible cable cooled by the Helium inlet at 4.5 K.

As any superconducting power line may be subject to an unexpected quench, the detector magnet protection circuit cannot pass through this flexible power line and a compact dump resistor has to be installed in the vicinity of the detector i.e. on the detector platform. The device will protect the superconducting coil from burning, in case of a quench, by extracting a substantial amount of magnetic energy from the coil itself and converting it into heat in the resistor bench. It is electrically connected in parallel to the coil current leads. Based on CERN experience dated from the nineteen seventies with the BEBC experiment and pursuing the study started at SLAC for SiD in 2007 (Ref. 13), a water based compact dump resistor is proposed (Ref. 14).

The cryogenics and vacuum plant is represented in Figure 3. On the surface of the experimental area, the He gas tank and liquid $N_2$ dewar ensure the storage of the coolants and the He gas compressor feeds the He liquefier (cold-box) that is located in the service area of the experiment cavern. Liquid He is then sent to a dewar close to the detector magnet and distributed via a valve box. A phase separator, located on top of the magnet, ensures the thermosiphon mode operation, i.e. the cooling He circulates without any active pump, the driving force being ensured by the difference in density between the liquid and the vapor. The connection between the helium liquefier and the dewar is via a flexible vacuum-insulated cryogenic transfer line that ensures the cooling of the magnet also during the detector movement from the garage to the beam position. The detailed engineering of such a line requires some R&D, namely for what concerns the minimum bending radius and the necessity of having an adequate slope between the equipment on the magnet and the cold-box. An alternative layout, with liquid He circulators is also envisaged. This solution, although more expensive, has some advantages compared with the thermosiphon cooling mode, in particular in a push-pull scenario, where the pressure drop inside the cryogenic transfer line is critical. The pumping system for the magnet cryostat is shown in Figure 3 as well. The fore-vacuum pumps, which are noisy and a source of vibrations, are located in the service area, far away from the beam position, while the high-vacuum pumps (usually stationary oil-diffusion pumps) are mounted on top of the magnet cryostat. The two systems are connected via a flexible rough vacuum pumping line that runs parallel to the cryogenic transfer line in a cable-chain above the detectors.

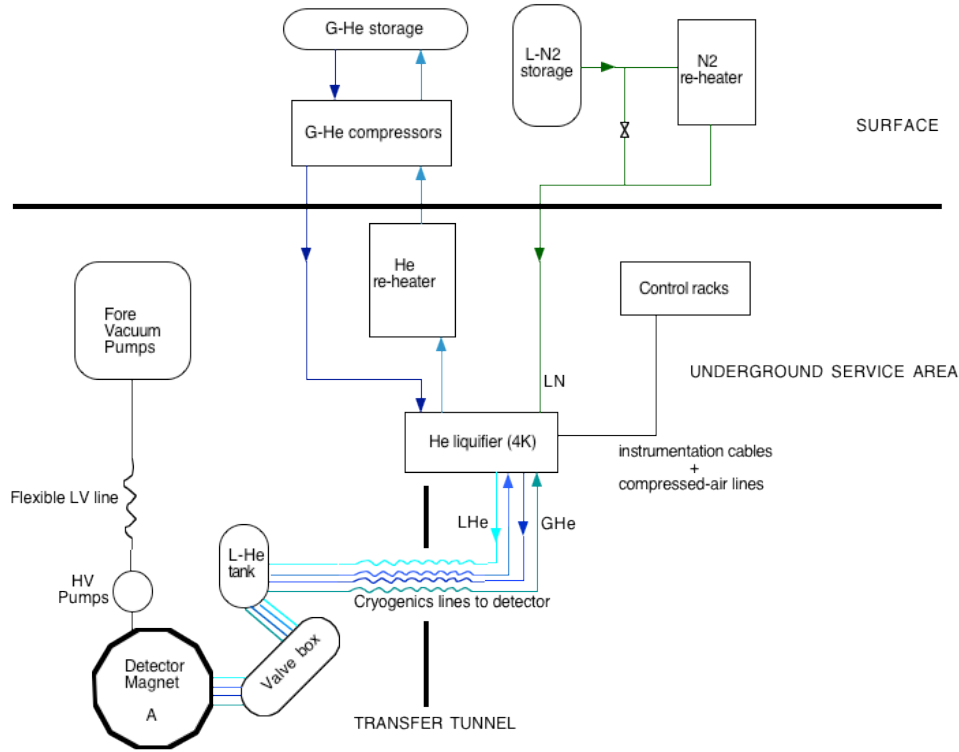

Figure 3. Schematics of the cryogenics and vacuum services.